# Virtual Private Environments for Multiphysics Code Validation on Computing Grids

Toan Nguyen°, Lizhe Wang°, Vittorio Selmin§

°Institut National de Recherche en Informatique et Automatique (INRIA Rhône-Alpes)
655, Av. de l'Europe, Montbonnot, F-38334 Saint-Ismier Cedex (France)
e-mail: Toan.Nguyen@inrialpes.fr
Web: http://www.inrialpes.fr/opale

§ALENIA Aeronautica S.p.A.
Corso Marche, 41, I-10146 Torino(Italy)

**Abstract:** Multiphysics simulation is the core of future engineering design in the aerospace industry. For this to become a production reality, quantum leap breakthroughs are to be achieved, concerning in particular model coupling, error correlations, alert definitions, best usage practices, code verification and code validation. Because problems that are expected to be orders of magnitude larger than current single discipline design are likely to be addressed, new computing technologies are required.

Among these technologies are parallel and distributed computing, in cluster and grid-based environments. It is clear that large PC-clusters and wide area grids are currently used for demanding numerical applications, e.g., nuclear and environmental simulation. It is not so clear however which approaches are currently the best for developing multiphysics simulation and validation environments. A first approach takes existing grid-based computing environments and deploys, tests and analyzes multiphysics codes. A second approach executes multiphysics codes to characterize grid-based environments for adequate architectural hardware and software.

We advocate in this paper the use of grid-based infrastructures that are designed for seamless approaches to the numerical expert users, i.e., the multiphysics applications designers. The approach is based on concepts defined by the HEAVEN* consortium. HEAVEN is a European scientific consortium including industrial partners from the aerospace and software industries, as well as academic research institutes.

The designers can define their own "virtual" computing environments by selecting the appropriate computing resources required, or reuse existing environments. The approach is generic by allowing various application domains to benefit from potential hardware and software resources located on remote computing facilities in a simple and intuitive way.

The computing resources are defined by services made available as sets of standardized interfaces performing specific tasks: application workflow, input data streams, output visualization tools, monitoring facilities, etc. Services can be composed and hierarchically defined. Transparent access to heterogeneous hardware and software operating systems is guaranteed. An aeroelasticity example is given.

**Key Words:** Multiphysics validation, Virtual problem-solving environments, Grid and cluster computing

---

## 1. Introduction

Single discipline simulation and optimization have made outstanding steps forward in the last two decades, including aerodynamics in hypersonic regimes, turbulence modeling, transonic regimes, etc. Significant efforts have been devoted to the validation of the corresponding codes, which is a prerequisite for their use in production environments. Powerful computing environments are also required for these codes to be run on realistic testcases. Current examples concerning 3D geometry optimization for wing and fuselage designs require tens of hours of CPU time on powerful computers.

It becomes clear that future applications of these methodologies in production environments will necessitate the coupling of several disciplines together. Significant progress has already been achieved in the aero-structure, aero-acoustics and electro-magnetics areas.
Two areas of interest raise new challenges here:

- what methodologies and approaches need be designed for multidiscipline simulation and optimization?
- what computer technologies and tools are best suited to support these methodologies?

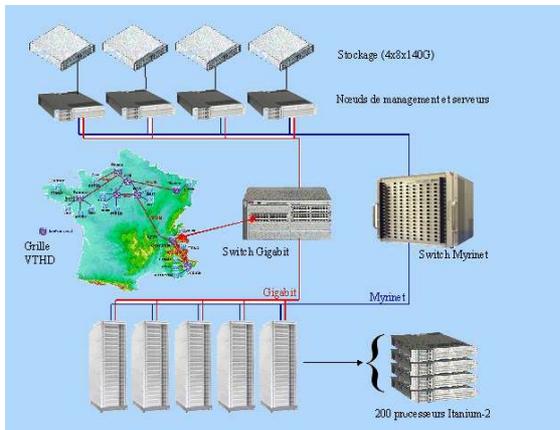

Figure 1. The VTHD grid infrastructure.

The approach emphasized here is the cross-leverage of parallel and distributed computing techniques, as supported by grid computing infrastructures, and adequate design and implementation techniques for numerical methods, e.g., domain decomposition, evolutionary algorithms, like genetic and game theory. It is indeed clear that the combined use of several nested levels of parallelism will provide efficient implementations of multidiscipline applications on parallel and distributed infrastructures. The French national VTHD grid infrastructure connecting a 100 Itanium2 PC-cluster at INRIA Rhône-Alpes is depicted by Figure 1.

With the help of some ideas proposed by world renown experts in their field [18], e.g., "virtual flight tests" and "integral operators", we propose in this paper some hints for the deployment of multiphysics code environments on future computing infrastructures.

Our vision is that the ever-growing complexity of computerized environments requires a parallel increase in usability and flexibility of their user interface. Far from the technology barrier hampering the wide dissemination of computing technology in developing countries, there are simultaneously "application pull" drivers that support the demand for ever increasing "technology push" drivers. This non-ending circle has to be made accessible to the application designers and to the end-users, which are not computer science experts.

A simple example is given by the rising tide of grid computing, i.e., the ability to use various computing resources and processors connected by wide-area networks as if it was a single computer for logging, resource reservation, accounting, security, etc. It is currently a technological burden to deploy, use and maintain such environments. The future lies in easy to use, transparent environments, in much the same way that the Internet can be used today by casual users and children altogether, totally unaware of the underlying technologies and infrastructures. An example of such interface developed in the CAST software project at INRIA Rhône-Alpes is given in Figure 2.

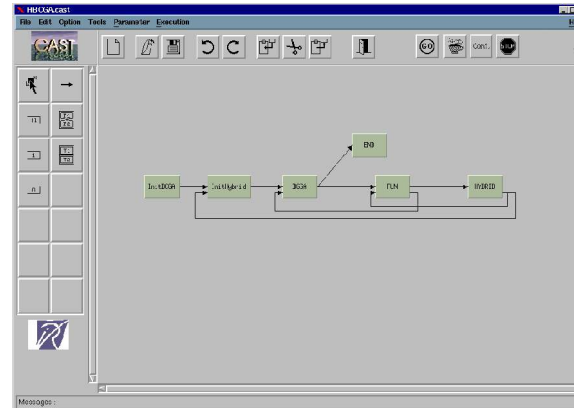

Figure 2. The CAST grid-computing interface.

Grid computing environments, a promising and maturing technology, is far to exhibit the Internet ease of use and friendliness, which took almost thirty years to achieve its current state-of-art. Because the premises of grid computing started in the mid-eighties, we can be quite optimistic…

However, several validated single discipline codes do not necessarily form a validated multidiscipline coupled-code, when linked together. Therefore, multi-scale modeling and simulations, multi-scale time stepping, and error correlations are but a few of the challenging issues raised by multidiscipline code validation.

Apart from coupling various disciplines, a net impact of distributed and parallel computing is here the scalability of the problems tackled. Indeed, current processing on large PC-clusters makes available solutions to problems that could not be designed and implemented five years ago.

Because the largest clusters include today thousands of processors[5], grid computing allows the deployment of multidiscipline applications on distributed resources that make such infrastructures a must….

The paper is organized as follows. Virtual private environments are introduced in Section 2. Their design and implementation on grid infrastructures is detailed in Section 3. Examples of multidiscipline applications in aerospace design are given in Section 4. Section 5 is a conclusion.

## 2. Virtual private environments

It is a common approach today for network design and deployment to share a common physical infrastructure among various logical and possibly overlapping layers of "virtual private networks" [10]. Such an approach bears a number of advantages among which are the ability to scale to the user communities needs, the security which is managed by the underlying infrastructure, the separation of domain addresses, etc.

"*A virtual private network (VPN) is a private data network that makes use of the public telecommunication infrastructure, maintaining privacy through the use of a tunneling protocol and security procedures. A virtual private network can be contrasted with a system of owned or leased lines that can only be used by one company. The main purpose of a VPN is to give the company the same capabilities as private leased lines at much lower cost by using the shared public infrastructure. Phone companies have provided private shared resources for voice messages for over a decade. A virtual private network makes it possible to have the same protected sharing of public resources for data. Companies today are looking at using a private virtual network for both extranets and wide-area intranets.*" In "VPN Technologies: definitions and requirements" (VPN Consortium white paper, July 2004, http://www.vpnc.org/vpn-technologies.html).

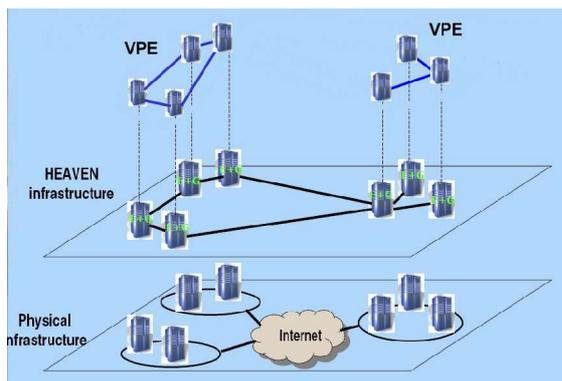

Figure 3. A virtual infrastructure.

This is similar to the "distributed virtualization" and "virtual testbeds" developed using overlay networks by the Planetlab consortium [19]. End-user services and "foundational sub-services" are deployed on virtual machines using disruptive technologies in synergistic testbeds and deployment platforms, which are "slices" of the underlying computing and network resources. But this is mainly a network operating system approach.

A different approach is used in HEAVEN for the design and deployment of "virtual private service environments" (VPE) on grid infrastructures [12].

Here, we focus on application development services rather than network operating system or end-users services. The goal is to share common computing resources, hardware and software, among various application groups for the secure, scalable and flexible design of application development services (Figure 3).

In contrast with current grid middleware, e.g., Globus, Unicore, the VPE do not support directly authentication, authorization, resource brokering and reservation[7, 21]. These are delegated to the grid middleware, which are specifically designed to do that. The added-value of VPE is precisely to mask the underlying middleware, in order to simplify access and use of grids. VPE enables the straightforward use of application codes without bothering about resource reservation. This is why we call the software layer in charge of VPE an "upperware". Its role is to generate, deploy and manage the VPEs.

VPE are sets of possibly overlapping high-level web services deployed by application designers to ease application design and deployment by the service providers. They bear some similarities with "virtual organizations" on grids [4]. But VPE can be implemented on computing environments without grid infrastructures, e.g., on networks of workstations, supercomputers, PCs, etc. They are a generic concept not necessitating grids. VPE are oriented to the application designers' communities. Their interface is a high-level graphic workflow definition and execution environment [14].

As such, VPE are sophisticated service layers building on the "upperware" that in turn relies and uses the middleware functionalities. VPE are not just another middleware.

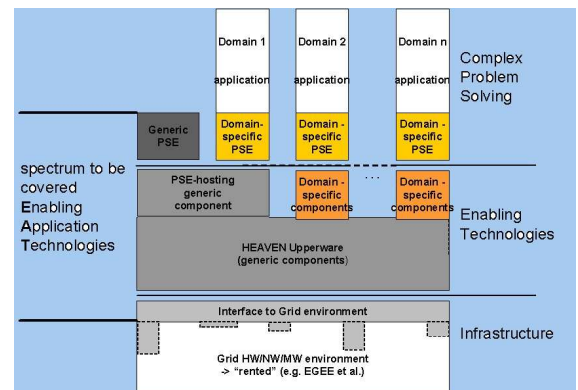

Figure 4. The HEAVEN architecture

## 3. Design and implementation

The upperware is a high-level software layer of sophisticated services. Its ultimate goal is to emulate the computing resources, hardware and software, and provide a hosting environment for the applications design and execution, based on powerful computational resources, e.g., grids. In contrast with other approaches specifically aimed at grids [10, 20] the hosting environment can here be any other infrastructure.

Conceptually, it is a hosting infrastructure that supports Virtual Private Environments emulating hardware and software resources. The environment supports applications that require specific resources. These resources range from computers to storage libraries and sensor devices or visualization and post-processing tools. Virtual environments are isolated from each other, securing the applications from unpredictable application behavior (Figure 4).

The underlying infrastructure ranges from mainframes to wide-area grids of PC-clusters. The current implementation relies on a testbed of several PC-clusters and workstations connected to a high-speed gigabits/sec network [22] (Figure 6).

The CAST [15] application management software is used on top of the Unicore middleware. The details of the testbed infrastructure are given in Table 1.

The user interface in CAST [13] is an intuitive graphic system which makes the supporting computer technology transparent (Figure 5).

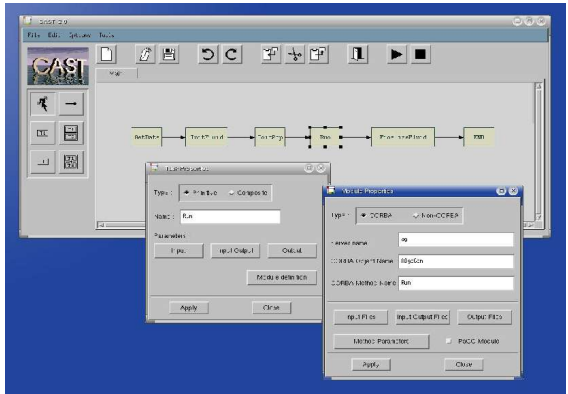

Figure 5. The CAST user interface.

The application components are linked by sequence, parallel and loop operators that from a high-level workflow. An example is detailed in Section 4. This workflow makes transparent all the technical details that are not strictly necessary to the end-users [11]. Components may be implemented in various programming languages, e.g., Fortran, C, C++, or Java. They may be parallel programs involving MPI statements. They may be compliant with CORBA or not, with J2EE containers or not. As such, software components of the application may be grid-aware or not.

Current developments of the HEAVEN upperware will make it compatible with WSDL [1] and WSRF [2] of Globus toolkit G.T.4[7]. This guarantees the compatibility with legacy software and future application software implemented on state-of-the-art middleware.

From a code validation perspective, connections to other devices are necessary. This includes flight-tests results, wind-tunnel experimental data. Ideally, these should be stored in databases for easy access through appropriate Web portals or servers. In this perspective, the connection of the VPE to these databases through the underlying infrastructure is straightforward: it is a matter of a few hours. However, performance tuning and assessment have to be established in order not to defeat the speed-up gained by using clusters and high-performance networks. In particular, an everlasting risk is the transmission delays related to the transfers of large volumes of data between application components. It requires the development and use of appropriate transfer protocols [6].

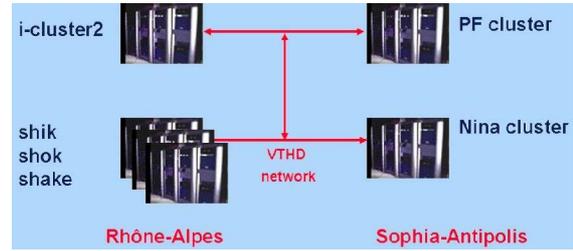

Figure 6. The testbed architecture.

| Resource | Hardware & OS | Software |
|---|---|---|
| **PF** @ Sophia Antipolis | Cluster: 19 nodes 100Mbps Fast-Ethernet 1 Node: 2× Pentium III @ 933Mhz Linux Kernel 2.4.2 & LSF | UNICORE server & CAST |
| **NINA** @ Sophia Antipolis | Cluster : 16 nodes 3 Gigabit-Ethernet 1 Node: 2× Xeon @ 2Ghz Linux Kernel 2.4.2 & LSF | |
| **i-cluster2** @ Rhône Alpes | Cluster : 100 nodes 1 Gigabit-Ethernet 1 Node: 2 × Itanium @ 900 MHz (64 bits) Linux Red Hat Advanced Server 3.0 | UNICORE server |
| **Shok** **Shik** **Shake** | Workstation: 1× Pentium III @ 1 GHz 100Mbps Fast-Ethernet Linux Fedora Core 2 | CAST, UNICORE client & server |

Table 1: Computing resources for the testbed.

## 4. Multidiscipline applications

An example of multidiscipline application is presented in this Section, concerning aeroelastic modeling and simulation. It was used in the Promuval project of the EC ("Prospective study on the state of the art of multidisciplinary modelling, simulation and validation in aeronautics" : http://www.cimne.upc.es/ PROMUVAL/). The goal was to test CAST as a software integration platform for multidiscipline validation in aeronautics. Major European aircraft manufacturers, together with research centers, where involved in PROMUVAL.

The example was provided by ALENIA Aeronautica (Italy). The goal is to study structural deformations of a medium-size airliner under specific aerodynamic conditions (Figure 7).

It includes the static and dynamic deformations of the wing structure under various load factors, corresponding to different cruise conditions at various speeds and altitudes, as well as pull-up and push-down maneuvers. The various testcases are described in Table 2.

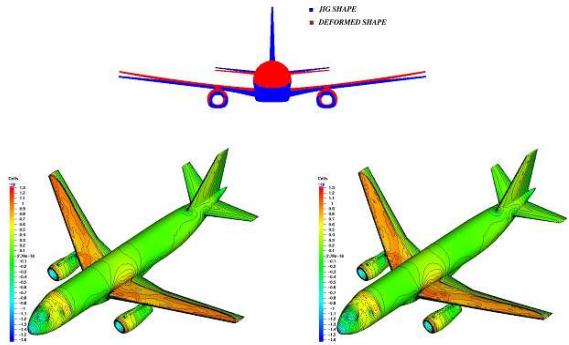

Figure 7. Aeroelastic testcase.

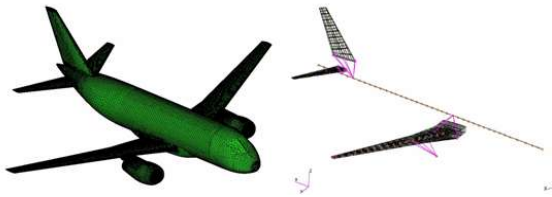

Table 2: Aero-stucture testcases.

The structural model is a simplified wingbox and tail assembly connected to a stick fuselage model and a boxed cell for the entire pylon+nacelle system (Figure 8).

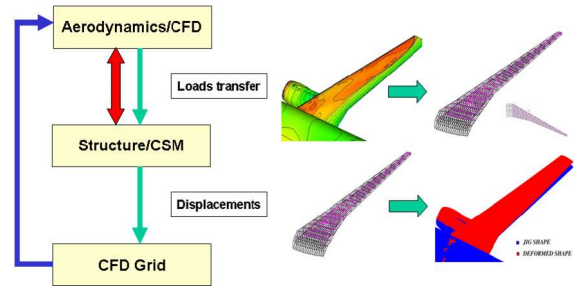

Figure 8. Stuctural model.

The interactions between the aerodynamics, structural mechanics and mesh generation are described in Figure 9.
The application workflow using the CAST integration platform is depicted Figure 10. It should be noted that although the interactions between the various application components are complex, because they involve a lot of parameters, the design of this workflow is simple. It is basically a loop involving the CFD, CSM and mesh generation/deformation components.

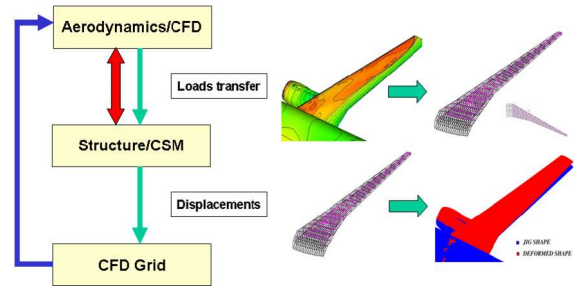

Figure 9. Model interactions.

This approach also masks the distribution and parallel implementation of the components. The application designers are the only persons to deal with these. This makes the end-users totally unaware of the underlying technical details.

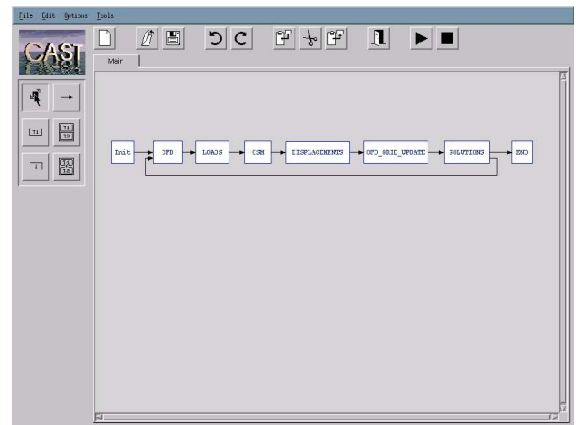

Figure 10. Aeroelasticity application workflow.

## 5. Conclusion

We present in this paper an approach for the deployment of multidiscipline applications on grid computing environments. It is based on a long term cooperation among aerospace and computer science experts from research labs and the industry.
An example provided by ALENIA Aeronautica (Italy) is detailed. It shows the transparent deployment of an aeroelasticity application on a grid infrastructure.
Although required by current multidiscipline applications for reasonable computing performance, the approach presented here requires also the connection to various databases for code validation. This is technically straightforward, but performance evaluation involving such databases remains to be assessed.
Multidiscipline simulation and optimization is the mainstream of research and development for the aerospace industry. Because it involves several disciplines (mathematical modeling, numeric optimization, computer science, e.g., distributed and parallel computing, grid and cluster infrastructures), its deployment in the production arena requires the extensive cooperation of various expertise. To achieve this goal however, a number of barriers remain. This includes the current complexity of computing technology, which hampers the easy use of sophisticated computing tools.

## Acknowledgements

The authors wish to thank the following persons for their strong support: Jean-Pierre Antikidis from CNES (French Space Agency), founder of the HEAVEN consortium, Alain Dervieux from INRIA, Jean-Antoine Desideri, head of the OPALE project at INRIA and Jacques Périaux (CIMNE, Barcelona, Spain, formerly at Direction de la Prospective, Dassault-Aviation, France).
This work was partly supported by the European Commission, projects DECISION, FLOWNET, INGENET and PROMUVAL.